\documentclass[aps,prl,reprint,groupedaddress,superscriptaddress,floatfix,nobibnotes]{revtex4-2}
\usepackage{amsmath}
\usepackage{mathtools}
\DeclarePairedDelimiter{\ket}{\lvert}{\rangle}
\usepackage{amssymb}
\usepackage{graphicx}
\usepackage{bm}
\usepackage[colorlinks=true,linkcolor=blue,citecolor=blue,urlcolor=blue]{hyperref}

\begin{document}

\title{Exact finite-sample and loss limits of squeezed-vacuum phase
estimation}

\author{Wangjun Lu}
\email{wjlu1227@zju.edu.cn}
\affiliation{School of Information Science and Engineering, Hunan Institute of Engineering, Xiangtan 411104, China}
\author{Qing Yu}
\email{yq@hnie.edu.cn}
\affiliation{School of Information Science and Engineering, Hunan Institute of Engineering, Xiangtan 411104, China}

\date{\today}

\begin{abstract}
We settle the dispute over squeezed-vacuum phase estimation by exact
computation of the finite-sample errors, resource-fair
scaling, phase window, and loss ceilings. A fixed overhead of ten repetitions
sets the attainable finite-sample frontier; the repetitions cancel
from the resource-fair bound, so Heisenberg scaling survives; the operating
window holds a repetition-set number of resolvable points and needs one bit of
branch prior; and fixed loss caps the per-photon advantage at a value set only
by transmissivity, already reached by the reported record.
\end{abstract}

\maketitle

\textit{Introduction.} Squeezed vacuum~\cite{caves1981} combined with homodyne detection has
recently been advanced as
a deterministic route to Heisenberg-limited phase estimation: two experiments,
one using direct homodyne detection of a single-mode squeezed
vacuum~\cite{nielsen2023} and one using a stimulated-emission nonlinear
(SU(1,1)) interferometer~\cite{yurke1986} with on--off single-photon
detectors~\cite{qin2023}, reported Fisher information per photon far above the
standard quantum limit (SQL) and claimed a scalable, unconditional
metrological advantage. Both claims have been challenged in a recent Comment
by Hradil and \v{R}eh\'a\v{c}ek~\cite{hradil2026}, who argue that
repetitions must be counted as resources, that the quantum Fisher
information (QFI) is not by itself an operational benchmark, and that the
advantage survives only in a phase window that shrinks with squeezing. A Reply~\cite{reply2026} defends the original
analysis.

The dispute has remained largely verbal because each side evaluates a
different idealization of the same protocol. The purpose of this Letter is to
compute, exactly and with no uncontrolled approximations, every quantity the
dispute touches, and thereby to adjudicate each claim on its own terms. The
protocol is simple enough that this program can be carried to completion: the
phase is encoded in the variance of a zero-mean Gaussian distribution, the
quadrature data admit a one-dimensional sufficient statistic, and the Bayesian
posterior can be evaluated without Gaussian or Laplacian approximations. Loss
can likewise be treated in closed form at both the homodyne-Fisher-information
and exact-QFI levels, verified against an independent density-matrix
calculation in the Fock basis.

Our conclusions are fourfold. (i) The Comment's central resource-counting
objection does not hold once the computation is exact: when the total photon
number $N=M\bar n$ is held
fixed, the number of repetitions $M$ cancels identically from the bound, so
the Heisenberg scaling survives, with a prefactor that includes the
finite-$M$ overhead. (ii) The Comment's window criticism is quantitatively
right and has a clean invariant form: the number of resolvable phase points
inside the operating window is $\sqrt{2M}$, independent of $\bar n$; the
advantage is therefore branch-conditional and narrow-range, not
``unconditional''. (iii) Neither side computed loss, and loss changes the
story qualitatively: any fixed transmissivity $\eta<1$ caps the per-photon advantage at
$1/[2(1-\eta)]$ (homodyne) or $1/(1-\eta)$ (optimal measurement), so the
quadratic per-photon growth ends at a plateau that depends only on $\eta$.
(iv) The two disputed experiments sit close to, and below, these ceilings.

\textit{Exact likelihood and finite-sample estimation.} The probe is a squeezed vacuum $\ket{\psi}=\hat S(r)\ket{0}$ with
$\bar n=\sinh^2 r$ and quadratures $\hat x=(\hat a+\hat a^\dagger)/\sqrt2$,
$\hat p=(\hat a-\hat a^\dagger)/(i\sqrt2)$, satisfying $[\hat x,\hat p]=i/2$. The phase shift
$\hat U(\phi)=e^{i\phi\hat a^\dagger \hat a}$ rotates the quadratures, so the homodyne
outcome $x$ at the optimal local oscillator angle is a zero-mean Gaussian
random variable with variance
\begin{equation}
V(\phi)=\tfrac12\bigl[\cosh 2r-\sinh 2r\,\cos 2\phi\bigr].
\label{eq:V}
\end{equation}
All phase information resides in the variance.

\begin{figure}[t]
\includegraphics[width=\linewidth]{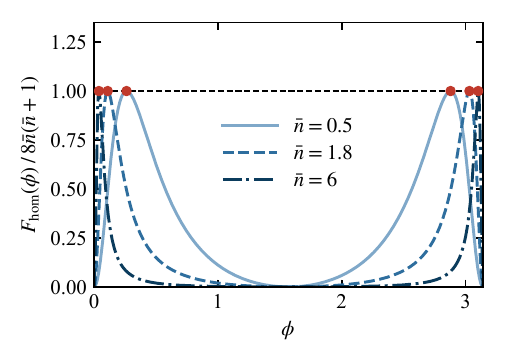}
\caption{(Color online) Homodyne Fisher information $F_{\rm cl}(\phi)$,
Eq.~\eqref{eq:Fcl}, in units of its maximum $8\bar n(\bar n+1)$, for
$\bar n=0.5$ (solid), $1.8$ (dashed), and $6$ (dash-dotted). Red dots mark
the optimal operating points $\phi^\ast$ of Eq.~\eqref{eq:phistar} and
their mirror images $\pi-\phi^\ast$. The information vanishes at $\phi=0$, peaks at
$\phi^\ast\simeq1/(4\bar n)$ [Eq.~\eqref{eq:phistar}], and the peak narrows
as $\bar n$ grows: the whole crest crowds against the zero-information
boundaries $\phi=0$ and $\phi=\pi$ at width $\sim1/\bar n$---a wave-crest
skin effect.
This profile generates the window and branch constraints
quantified in the sections that follow.}
\label{fig:fisher}
\end{figure}

The parameter is a dynamical phase, not a Hamiltonian coupling: it enters
through $\hat U(\phi)$ alone, so $\partial_\phi$ acts on the state family and not
on any operator, and $F$ is invariant under mode-basis relabelings,
local-oscillator phase choices, and gauge-like reparametrizations of the field.
No ambiguity of the kind that afflicts estimates of a coupling strength arises;
the only accounting freedoms left---the resource unit and the phase
prior---are fixed explicitly below. For any zero-mean Gaussian
distribution the classical Fisher information of a single sample is
\begin{equation}
F_{\rm cl}(\phi)=\frac{(\partial_\phi V)^2}{2V^2},
\label{eq:Fcl}
\end{equation}
which is maximized at $u^\ast=\tanh 2r$ with $u=\cos2\phi$, i.e.\ at the
operating point
\begin{equation}
\phi^\ast=\tfrac12\arccos(\tanh 2r)\simeq\frac{1}{4\bar n}\quad(\bar n\gg1),
\label{eq:phistar}
\end{equation}
where
\begin{equation}
F_{\rm max}=2\sinh^2 2r=8\bar n(\bar n+1)=4\,{\rm Var}(\hat n),
\label{eq:Fmax}
\end{equation}
which equals the QFI of the pure probe. Three independent evaluations---the
analytic form~\eqref{eq:Fcl}, direct numerical integration of
$p(x|\phi)(\partial_\phi\ln p)^2$, and a Fock-space calculation of
$4\,{\rm Var}(\hat n)$---agree to machine precision. Note that $\phi^\ast\neq\pi/4$ and $F_{\rm cl}(0)=0$: the operating point
adjoins a zero-information point, which underlies everything that follows.
Figure~\ref{fig:fisher} plots this profile for several $\bar n$. The
geometry is a wave-crest skin effect: as $\bar n$ grows, the entire
information crest piles up against the zero-information boundaries
$\phi=0$ and $\phi=\pi$, with crest position $\phi^\ast\propto1/\bar n$ and
width $\sim1/(2\bar n)$; the boundary value stays pinned at zero, so this is
a self-similar crest collapse (rising as $\phi^2$, decaying as $\phi^{-2}$)
rather than a surface current.

Equation~\eqref{eq:Fmax} is the resource statement in its standard
form~\cite{giovannetti2011}: for unitary encoding, with $\sigma$ the phase
uncertainty,
$\sigma\ge1/(2\sqrt{M}\,\Delta\hat H)$ and $F_Q=4\Delta\hat H^2$ for a pure
probe~\cite{braunstein1994}, with $\hat H=\hat n$ and
$\Delta\hat n=\sqrt{2\bar n(\bar n+1)}\simeq\sqrt2\,\bar n$. The squeezed
vacuum thus reaches Heisenberg scaling by making the spread of the generator linear in
the photon number, not by any non-extensive generator: the $\bar n^2$ growth of
$F$ is the square of a linear $\Delta\hat n$, and nothing beyond the
Heisenberg bound survives on the $\Delta\hat n$ axis. It is this identification
that loss breaks, as we show below.

\begin{figure}[t]
\includegraphics[width=\linewidth]{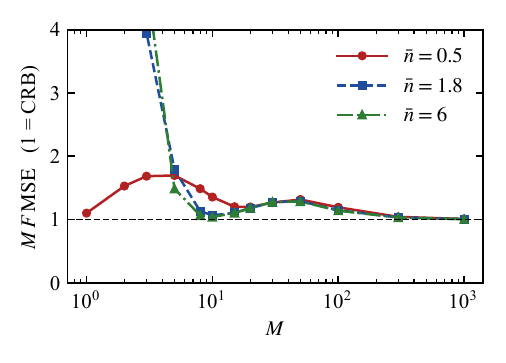}
\caption{(Color online) $M\,F_{\rm max}\times{\rm MSE}$ versus repetitions
$M$ (exact Bayesian computation) for $\bar n=0.5$ (circles, solid),
$1.8$ (squares, dashed), and $6$ (triangles, dash-dotted); unity means the
CRB is attained. The minimum sits near $M^\ast\simeq10$ for every $\bar n$.}
\label{fig:conv}
\end{figure}

For $M$ independent homodyne outcomes the joint likelihood depends on the data
only through the sufficient statistic $S=\sum_i x_i^2$,
\begin{equation}
p(\{x\}|\phi)\propto V(\phi)^{-M/2}\,
e^{-S/2V(\phi)},\qquad
S/V(\phi^\ast)\sim\chi_M^2,
\label{eq:lik}
\end{equation}
so the exact finite-$M$ Bayesian posterior is one-dimensional and can be
evaluated on a grid without approximation. Using the posterior mean as
the estimator, Fig.~\ref{fig:conv} shows the product $M\,F_{\max}\,{\rm MSE}$,
which equals unity when the Cram\'er--Rao bound (CRB) is attained.

Two exact results follow. First, the CRB is approached once $M$ is of order
ten: for $M\ge10$ the exact excess $M\,F_{\rm max}\,{\rm MSE}-1$ stays bounded,
$0.01$--$0.36$, over $\bar n\in[0.5,1000]$ (largest at small $\bar n$,
reflecting the heavy-tailed posterior mean), and the threshold
$M^\ast\simeq10$ does not grow with $\bar n$. The reason is transparent from
Eq.~\eqref{eq:lik}: the relative fluctuation of $S$ is governed by the
$\chi_M^2$ distribution, whose width scales as $1/\sqrt M$ independently of
$\bar n$, while the phase curvature is absorbed in $F$. Second, at $M=1$ the
MSE exceeds the CRB by a factor of $8.3$ ($\bar n=1.8$) to $145$ ($\bar n=11$):
quantifying the Comment's objection that a single-shot Fisher information
is not an operational benchmark; the objection is real but bounded---the
excess drops below $36\%$ by $M\simeq10$ and never grows with the resource.

\begin{figure}[t]
\includegraphics[width=\linewidth]{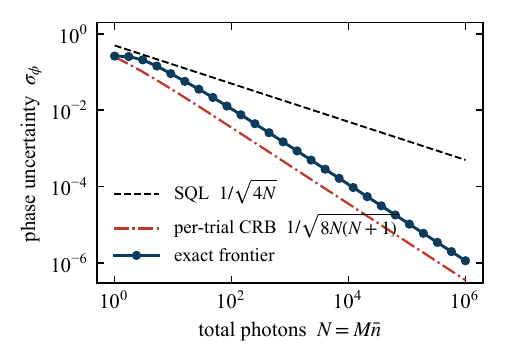}
\caption{(Color online) Resource-fair frontier: exact Bayesian computation of
the attainable phase uncertainty $\sigma$ versus total photon number
$N=M\bar n$ (symbols), compared with the SQL $1/\sqrt{4N}$ (dashed) and the
per-trial CRB $1/\sqrt{8N(N+1)}$ (dash-dotted). The frontier follows the CRB
over six decades in $N$ and lies $44$ times below the SQL at $N=10^4$.}
\label{fig:frontier}
\end{figure}

\textit{Resource accounting: the repetitions cancel.} The resource-fair comparison fixes the total photon number $N=M\bar n$ passing
through the phase element. Whenever the CRB is valid ($M\gtrsim M^\ast$),
\begin{equation}
{\rm Var}=\frac{1}{M F_{\rm max}}
=\frac{1}{8N(\bar n+1)}
\;\longrightarrow\;
\frac{1}{8N(N/M^\ast+1)}.
\label{eq:cancellation}
\end{equation}
Adding repetitions with fewer photons each, or fewer repetitions with more
photons each, yields the same total information---the algebraic refutation of
the claim that counting repetitions restores the classical $1/N$ scaling. The only residue of
finite statistics is the fixed overhead $M^\ast\simeq10$, which converts the
single-shot value $1/\sqrt{8N(N+1)}$ into the attainable frontier
\begin{equation}
{\rm Var}_{\rm min}\simeq\frac{1.30}{N^2},
\qquad
\sigma\simeq\frac{1.14}{N},
\label{eq:frontier}
\end{equation}
where the arrow in Eq.~\eqref{eq:cancellation} sets $M=M^\ast$: because
$F_{\max}$ grows with $\bar n=N/M$, the optimum sits at the smallest
admissible repetition number. Exact Bayesian evaluation at $\bar n=1000$, $M=10$,
$N=10^4$ gives MSE $=1.30\times10^{-8}$, i.e.\ $M\,F_{\rm max}\,{\rm MSE}=1.04$, within
$4\%$ of $1/(M^\ast F_{\max})=1.249\times10^{-8}$---the residual being the
finite-sample excess of Fig.~\ref{fig:conv}---and attaining the
frontier~\eqref{eq:frontier}. At $N=10^4$ this is $44$ times below
the SQL $\sigma_{\rm SQL}=1/(2\sqrt N)$; Figure~\ref{fig:frontier} plots the
frontier over six decades in $N$. The Heisenberg
scaling claimed in Refs.~\cite{nielsen2023,qin2023} is therefore real, but the
precise statement of it carries the prefactor $1.14$ and the constraint
$M\gtrsim10$---neither of which appears in the original claims, and neither of
which destroys the scaling as the Comment argues.

\begin{figure}[t]
\includegraphics[width=\linewidth]{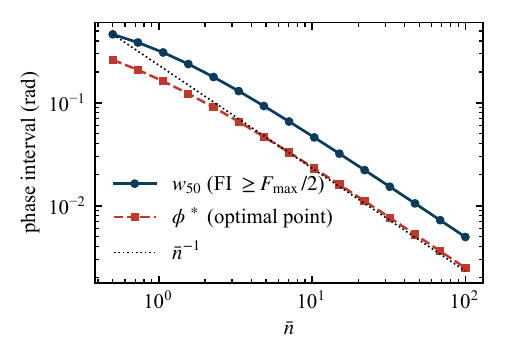}
\caption{(Color online) The shrinking window: numerically extracted full width
$w_{50}$ of $F_{\rm cl}(\phi)$ (at $F\ge F_{\max}/2$) and the operating point
$\phi^\ast$, versus $\bar n$ (log--log). Both follow the $\bar n^{-1}$ power
law (dotted guide), confirming Eq.~\eqref{eq:window}; the number of resolvable
points inside the window, Eq.~\eqref{eq:capacity}, is set by $M$ alone.}
\label{fig:window}
\end{figure}

\textit{Window, capacity, and the one-bit prior.} The Comment's strongest point is that the advantage occupies a shrinking
phase window. Equation~\eqref{eq:Fcl} vanishes at $\phi=0$, and the
full width at half maximum of $F_{\rm cl}(\phi)$ is
\begin{equation}
w_{50}=2\phi^\ast\simeq\frac{1}{2\bar n},
\label{eq:window}
\end{equation}
verified numerically (ratio converging to $2.000$). Combining
Eqs.~\eqref{eq:phistar}--\eqref{eq:window}, the number of resolvable phase
points inside the window is
\begin{equation}
\frac{w_{50}}{\sigma}=\sqrt{2M},
\label{eq:capacity}
\end{equation}
independent of $\bar n$: squeezing sharpens the local readout
($\sigma\propto1/\bar n\sqrt M$) and narrows the window ($w_{50}\propto1/\bar n$)
in equal measure. This is the precise
version of the Comment's ``stopped clock'' metaphor: the dial shrinks with
the precision, and its information capacity is set by $M$ alone.
Figure~\ref{fig:window} verifies both power laws numerically.

The variance \eqref{eq:V} is $\pi$-periodic and mirror-symmetric,
$\phi\leftrightarrow\pi-\phi$, so the two branches are indistinguishable
without prior information. With a flat prior on $[0,\pi)$ the posterior is
exactly bimodal and the rms error worsens at $M=100$ by factors of $88$ ($\bar n=1.8$) to
$480$ ($\bar n=11$) relative to the branch-resolved case. One bit of prior
information---which the experiments supply by construction---restores the
performance; since it does not scale with $\bar n$ it affects constants, not
scaling, but it belongs on the resource list: the advantage is
branch-conditional and narrow-range, not ``unconditional''.

\textit{Loss: closed-form ceilings neither side computed.} In metrology, loss is the standard route by which quantum enhancements are
degraded~\cite{demkowicz2012}; here it is the ingredient neither
side computed.
A loss channel with transmissivity $\eta$ maps the zero-mean Gaussian probe to
another zero-mean Gaussian state, so Eq.~\eqref{eq:Fcl} still applies with
\begin{align}
V_\eta(\phi)&=\tfrac12\bigl[A-B\cos2\phi\bigr],\nonumber\\
A&=1+\eta(\cosh2r-1),\qquad B=\eta\sinh2r.
\label{eq:Vloss}
\end{align}
Maximizing Eq.~\eqref{eq:Fcl} with Eq.~\eqref{eq:Vloss} gives the
homodyne Fisher information
\begin{equation}
F_{\rm hom}(\eta)=\frac{2B^2}{A^2-B^2},
\label{eq:Fhom}
\end{equation}
while the exact QFI~\eqref{eq:FQ} of the mixed state, obtained from the covariance matrix
with the purity correction and verified against an independent SLD (symmetric
logarithmic derivative) computation in the Fock basis (agreement to machine precision for $\bar n\le2$ and to
$1.3\times10^{-4}$ at $\bar n=4$, limited only by Fock truncation), is
\begin{equation}
F_Q(\eta)=\frac{4B^2}{1+A^2-B^2}\;\ge\;F_{\rm hom}.
\label{eq:FQ}
\end{equation}
At $\eta=1$ both reduce to $2\sinh^2 2r$; for $\eta<1$
Eq.~\eqref{eq:FQ} exceeds Eq.~\eqref{eq:Fhom}, with the excess approaching $100\%$ in the large-$\bar n$ limit;
homodyne reaches only $57\%$ of the exact QFI at $\eta=1/2$, $\bar n=6$
(Table~\ref{tab:loss})---a gap familiar from lossy Gaussian
metrology~\cite{monras2007,pinel2013}---so the assumption that homodyne
detection is optimal holds only for a pure probe.

\begin{table}[b]
\caption{Homodyne Fisher information $F_{\rm hom}(\phi^\ast)$ versus exact
QFI $F_Q$ of the lossy probe (transmissivity $\eta$, input $\bar n$). The two
coincide only at $\eta=1$.}
\label{tab:loss}
\begin{ruledtabular}
\begin{tabular}{ccccc}
$\eta$ & $\bar n$ & $F_{\rm hom}(\phi^\ast)$ & $F_Q$ (SLD) & $F_{\rm hom}/F_Q$\\
\colrule
1.00 & 1.8 & 40.320 & 40.320 & 1.000\\
0.89 & 1.8 & 18.733 & 23.615 & 0.793\\
0.89 & 6.0 & 79.456 & 122.203 & 0.650\\
0.74 & 6.0 & 32.753 & 55.552 & 0.590\\
0.50 & 6.0 & 12.000 & 20.985 & 0.572\\
\end{tabular}
\end{ruledtabular}
\end{table}

\begin{figure}[t]
\includegraphics[width=\linewidth]{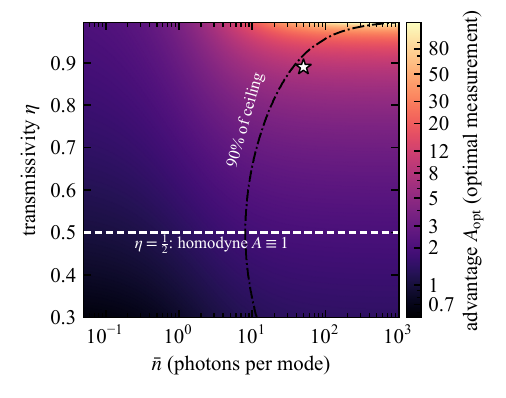}
\caption{(Color online) Per-photon advantage
$A_{\rm opt}=F_Q/(4\eta\bar n)$ over a coherent probe at the same photon number
and loss, over the plane of mean photon number $\bar n$ and transmissivity
$\eta$ (color scale). $A_{\rm opt}$ is a smooth function of both variables,
rendered without contour levels, and grows monotonically with $\bar n$
toward the ceiling $1/(1-\eta)$ of Eq.~\eqref{eq:ceiling}.
Dash-dotted curve: 90\% of the ceiling reached. White dashed line:
$\eta=1/2$, where the homodyne advantage is exactly unity for all $\bar n$
($A_{\rm opt}=2$). Star: reported record of Ref.~\cite{nielsen2023}
($\eta=0.89$), already on the plateau.}
\label{fig:ceiling}
\end{figure}

For $\bar n\gg1$ at fixed $\eta$, $A^2-B^2\simeq4\eta(1-\eta)\bar n$ and the
quadratic growth terminates:
\begin{equation}
F_{\rm hom}\to\frac{2\eta\,\bar n}{1-\eta},
\qquad
F_Q\to\frac{4\eta\,\bar n}{1-\eta}.
\label{eq:plateau}
\end{equation}
Against the coherent-state baseline $F_{\rm coh}=4\eta\bar n$ at the same $\bar n$ and
loss, the plateaus~\eqref{eq:plateau} hard-cap the per-photon advantages
at
\begin{equation}
A_{\rm hom}^{\max}=\frac{1}{2(1-\eta)},
\qquad
A_{\rm opt}^{\max}=\frac{1}{1-\eta},
\label{eq:ceiling}
\end{equation}
independent of the probe; Fig.~\ref{fig:ceiling} maps $A_{\rm opt}$ over
the whole plane and shows the saturation geometry at a glance. At $\eta=1/2$ the homodyne
advantage is \emph{exactly} unity for all $\bar n$: substituting
$\eta=1/2$ into Eq.~\eqref{eq:Fhom} gives $F_{\rm hom}=2\bar n=4\eta\bar n$
identically, a closed-form identity we verify numerically over
$\bar n\in[0.78,300]$. Squeezing below $\eta=1/2$ is strictly counterproductive
for homodyne detection, while optimal measurement retains a factor-two
advantage: the useful threshold is measurement-dependent.

\begin{figure}[t]
\includegraphics[width=\linewidth]{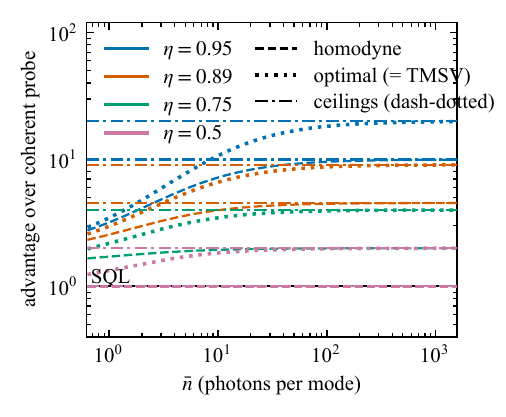}
\caption{(Color online) Per-photon advantage over a coherent probe at the
same photon number and loss, versus $\bar n$ (per mode). Colors distinguish
the transmissivity: blue $\eta=0.95$, orange $\eta=0.89$, green
$\eta=0.75$, purple $\eta=0.5$. Line styles distinguish the measurement:
dashed, single-mode homodyne; dotted, optimal measurement, where the
single-mode and two-mode (TMSV) curves coincide exactly
[Eq.~\eqref{eq:tmsv}]. Dash-dotted horizontal lines in the same colors: the
ceilings of
Eq.~\eqref{eq:ceiling} for each $\eta$---upper $1/(1-\eta)$ (optimal),
lower $1/[2(1-\eta)]$ (homodyne); solid black line: the SQL (unity). For
$\eta\le0.75$ each ceiling coincides with the asymptote of its curve, so
dash-dotted line and curve overprint there.
The $\eta=0.5$ homodyne curve sits exactly at unity for all $\bar n$---the
threshold identity below Eq.~\eqref{eq:ceiling}.}
\label{fig:universal}
\end{figure}

The same construction settles the comparison between the two disputed
protocols. With symmetric modes $\hat c_\pm=(\hat a\pm\hat b)/\sqrt2$ the two-mode squeezer
factorizes, $\hat S_{\rm TM}(r)=\hat S_{c_+}(r)\hat S_{c_-}(-r)$: the two-mode squeezed vacuum
(TMSV) is a product of two single-mode squeezed vacua with the same $r$, equal loss on
$\hat a$ and $\hat b$ acts independently on each $\hat c_\pm$, and the generator
$\hat N_a+\hat N_b=\hat N_{c_+}+\hat N_{c_-}$ is additive. Hence
\begin{equation}
F_Q^{\rm TMSV}(\eta,r)=2F_Q(\eta,r),\qquad N=2\bar n.
\label{eq:tmsv}
\end{equation}
Eq.~\eqref{eq:tmsv} reproduces $4\sinh^22r=4N(N+2)$ at
$\eta=1$ (with $N=2\bar n$ the total photon number)---the value quoted in
Ref.~\cite{qin2023}---and, dividing by the
doubled photon number,
gives \emph{exactly} the same per-photon advantage curve as the single-mode
protocol, not merely the same asymptote; Figure~\ref{fig:universal} displays
the resulting ceiling structure. The two schemes are informationally
equivalent per photon: they differ only in how loss is placed. An exact
Fock-space SLD calculation shows that placing the loss in one arm changes the
QFI by less than $13\%$ relative to equal splitting of the same
\emph{total} transmissivity ($\bar n=1$--$4$), so Eq.~\eqref{eq:ceiling}
applies to either loss placement.

\textit{The reported records sit on the plateau.} The single-mode experiment reports $F=15.8(6)~{\rm rad}^{-2}$ per photon at
a total loss of $11\%$~\cite{nielsen2023}. The homodyne ceiling at $\eta=0.89$
is $2\eta/(1-\eta)=16.2~{\rm rad}^{-2}$ per photon: the reported record is at
$97\%$ of it, inside the error bar, and lies on the plateau of
Eq.~\eqref{eq:ceiling} under either reading of the normalization (largest
probed photon number with $\eta=0.89$, or $\bar n\simeq1.8$ with effective
$\eta\simeq0.955$). The
nonlinear-interferometer experiment reports a $5.8(1)$-fold
enhancement~\cite{qin2023}, corresponding to the ceiling for an internal loss
of $9\%$--$17\%$ depending on whether its on--off detection remains
QFI-optimal under loss; either way, the enhancement cannot grow with squeezing.
The advantage in both experiments is real but already saturated, and
``Heisenberg-limited scaling'' is a statement about the lossless model, not
about the experiments.

\textit{Conclusions.} We have given an exact, closed-form account of squeezed-vacuum phase
estimation at every level the recent dispute appeals to. The repetitions cancel from the resource-fair bound, so the Comment's
central objection does not survive the exact computation; the window and
branch criticisms are promoted to
invariants, Eqs.~\eqref{eq:window}--\eqref{eq:capacity} and the one-bit prior;
and loss imposes ceilings \eqref{eq:ceiling} that the reported records already
approach. The framework transfers verbatim to any protocol with the parameter encoded
in a covariance matrix, including cavity-optomechanical coupling estimates,
where the same finite-$M$, window, branch, and loss analyses must accompany
any QFI-based claim.

\begin{acknowledgments}
This work was supported by the National Natural Science Foundation of China
(Grant No.~12205092), the Hunan Provincial Natural Science Foundation
(Grant No.~2026JJ50100), and the Scientific Research Fund of the Hunan
Provincial Education Department (Grant No.~24B0670).
\end{acknowledgments}

\end{document}